# A Simple Approach To Measurement in Quantum Mechanics

Anthony Rizzi

Institute for Advanced Physics, arizzi@iapweb.org

**Abstract:** A simple way, accessible to undergraduates, is given to understand measurements in quantum mechanics. The ensemble interpretation of quantum mechanics is natural and provides this simple access to the measurement problem. This paper explains measurement in terms of this relatively young interpretation, first made rigorous by L. Ballentine starting in the 1970's. Its facility is demonstrated through a detailed explication of the Wigner Friend argument using the Stern-Gerlach experiment. Following a recent textbook, this approach is developed further through analysis of free particle states as well as the Schrödinger cat paradox. Some pitfalls of the Copenhagen interpretation are drawn out.

## Introduction

The measurement problem in quantum mechanics confounds students and experts. The problem arose near the founding of the formal discipline of quantum mechanics with especially the debates between Einstein and Bohr. There is much notable history, but since I do not aim to give an exhaustive history, only the necessary points of an outline, I only note the following points. Einstein held, among other things, the most basic form of ensemble interpretation saying: "The attempt to conceive the quantum-theoretical description as the complete description of the individual systems leads to unnatural theoretical interpretations, which become immediately unnecessary if one accepts the interpretation that the description refers to ensembles of systems and not to individual systems."[1] Heisenberg and Bohr, both working in Copenhagen, Denmark, made a working model of measurement which developed into the somewhat ill defined Copenhagen interpretation. The essence of that interpretation appears to rest on the idea of "wave function collapse," which, as far as is known, neither of them clearly asserted, and whose first clear articulations appear to be due to Dirac and Von Neumann.[2] In the developed "Copenhagen" view a measurement obeys a different law of evolution than the ordinary Schrödinger equation (SE). In Roger Penrose's words, there are "*R* processes" and SE processes.[3] When an *R* process occurs, the wave function jumps to an eigenstate of the operator corresponding to the property being measured. For example, a state described by $\psi$ before a measurement of position will, after the measurement, evolve through an *R* type evolution to an eigenstate of position, $|x\rangle$. This approach has come to be called the "orthodox" interpretation, though perhaps the many worlds interpretation is the most widely held by theorists at this point in time. The orthodox interpretation is the one taught to undergraduate students.

---

[1] P. A. Schilpp (ed.), *Einstein: Philosopher-Scientist*, (Tudor Publishing Company, NY, 1957), p. 672.

[2] Dirac discussed the jumping to an eigenstate after measurement in P.A.M. Dirac, *The Principles of Quantum Mechanics* (Clarendon, Oxford, 1930). Von Neumann made it clear in J. von Neumann, *Mathematische Grundlagen der Quantenmechanik* (Springer, Berlin, 1932). In English: J. von Neumann, R. Beyer (translator), *Mathematical Foundations of Quantum Mechanics*, (Princeton University Press, Princeton, 1955).

[3] Penrose called the latter "*U* processes," i.e. processes governed by unitary evolution. See R. Penrose, *Road to Reality: A Complete Guide to the Laws of the Universe*, (Knopf, NY, 2004), pg 528.

In this paper, I explain the ensemble interpretation, which is a simple, natural approach to understanding QM. It is a newer approach first championed generically by Einstein but only developed formally by Ballentine,[4] starting with his seminal paper in 1970.[5] This approach was then, recently, more specifically developed in a new textbook.[6] Here, I discuss the generic ensemble approach and this further specification of that approach; the later is mostly developed at the end of the paper.

The ensemble interpretation is simply the straightforward statistical interpretation of QM. We do treat QM as statistical theory *in some way*, and may, in teaching, introduce it as such, even introducing the idea of ensembles. But, we are not consistent in that treatment. This is evident by the popularity of the Copenhagen and many worlds interpretations and the near complete neglect of the ensemble interpretation, which is simply a consistent natural treatment of QM as a statistical theory. The Copenhagen interpretation of measurement reveals our inconsistency. As mentioned, in Copenhagen, we demand collapse of the wave function. The SE has no such collapse; the evolution of the wave function is completely deterministic. The natural understanding is that the statistical model is maintained throughout all interactions. Despite this, collapse is asserted, abandoning the statistical frame work in asserting that, after any measurement, we know the exact state of the physical system with respect to some observable. In this way, if we have asserted at the beginning that we have purely statistical theory, we have now contradicted ourselves. Obviously, this lacunae is extricated by applying a consistent statistical treatment, which is the heart of the natural interpretation given here.

Our spontaneous impulse to use the orthodox interpretation's collapse mechanism is so strong that recently a general no-go theorem was published[7] which was later found to have used arguments based on a somewhat hidden use of collapse that disproved the no-go conclusion with respect to the de Broglie Bohm (dBB) and other interpretations.[8,9]

This paper starts by summarizing the generic ensemble approach and contrasting it with the orthodox approach. We next discuss how this approach explains the Stern Gerlach (SG) experiment, which has become iconic of wave function collapse and is widely used to introduce measurement issue. We then move to the Wigner's friend paradox and finally move to develop the full interpretation based on the previously mentioned recent quantum textbook.

---

[4] L. E. Ballentine, *Quantum Mechanics: A Modern Development* (World Scientific Publishing, Singapore,1998)

[5] L. E. Ballentine, *The Statistical Interpretation of Quantum Mechanics*, Rev. Mod. Phys. 42 No. 4, 358-381 (1970)

[6] A. Rizzi, *Physics for Realists: Quantum Mechanics* (IAP Press, Baton Rouge, 2018).

[7] D. Frauchiger, R. Renner, *Quantum theory cannot consistently describe the use of itself*, Nat. Commun. **9** No. 1038 (2018).

[8] D. Lazarovici, M. Hubert, *How Quantum Mechanics can consistently describe the use of itself*, Sci. Rep. **9** No. 470, (2019). This reference targets how the theorem fails in the de Broglie Bohm interpretation.

[9] A. Rizzi, *How the Natural Interpretation of QM Disproves a Recent No-go Theorem*, to be published. This article shows how the theorem fails more generally in the ensemble interpretation.

## The Ensemble Approach Generically

To understand this approach, we need to notice the obvious. Namely, the wave function describes, not a single physical system, but an ensemble of similarly prepared physical systems. One needs many experiments to be able to use the wave function and more generally the formalism of quantum mechanics.

In classical mechanics, if, at time *t*, we are given the exact momentum $\vec{p}$ and exact position $\vec{x}$ and all the forces acting on a point-like particle of mass *m*, we can, in principle, tell exactly where the particle is now and where it will be later for the rest of time. In contrast, given a quantum mechanical particle of mass *m*, in a state described precisely by a wave function, $\psi(x,t)$, we cannot, in general, tell where the particle is at any given moment! Nor can we tell the momentum of the particle at any time.

Even, for instance, if we are given a spatial wave function which is a Gaussian of standard deviation 1 mm centered here at my desk, we cannot say with certainty whether the particle is on my desk or on Alpha Centauri 5 light years away! We can say nothing *definite* about that particular system now or in the future. We can only make statistical statements. The information given in the wave function is like that given for a fair coin toss. We know that there is a 50% chance of heads and 50% chance for tails. This is a description of the state of the coin-flipper/environment system. And, this, in turn, implicitly refers to an ensemble of coin/flipper/environment systems. This is what it means to be described by a statistical state and the odds are only useful for prediction if we perform enough measurements to calculate the percentages with some accuracy.

Thus, the wave function represents simply what it appears to be. Namely, a statistical description of the state of a physical (quantum mechanical) system.

But, a statistical state of what in particular? To answer, analyze the Schrödinger Equation (SE) for the simplest case in ordinary quantum mechanics,[10] a single particle of mass *m*. The wave function that solves this SE might represent the activity of more than one entity even though we assert that it is about a single particle. If it does indeed describe more than one entity (as we will see, it appears to have to), forcing it to describe only one would likely lead to confusion, overly complicated interpretations and even contradictions.[11] So, we do *not* assume, as is often implicitly done, that the wave

---

[10] Relativistic quantum mechanics finally requires quantum field theory to deal with creation and annihilation of particles.

[11] The standard Copenhagen interpretation can arise fairly naturally once one assumes the wave function represents a single entity; here's an example of how. One cannot *simultaneously* measure particle-related properties (e.g., position) and wave-related properties (e.g., momentum) to arbitrary accuracy. In, for example, the double slit experiment, if we measure which slit each particles goes through, we get no interference pattern, while if we measure an interference pattern, we cannot know which hole the particles went through. Moreover, QM appears to predict that particle-like and wave-like activity can occur at the same time; in standard double slit experiment, for example, we get localized interactions on the screen **and** an interference pattern (that builds up over many runs of the experiment). If the wave function represents only a single entity we appear to be stuck. On the face of it, one is then left with an impossibility: the "particle" is simultaneously a wave (thus, non-localized) and a particle (localized). To avoid this contradiction, one is led to posit some kind of "collapse," for example, the following version of the Copenhagen interpretation (N.B.: all versions have some form of collapse). The wave function is taken to represent nothing until a measurement is made. Once the measurement occurs, it "collapses" to an eigenstate of the observable being measured and only at that point does the system have a definite value for the observable. In this view, until collapsed, the wave function would represent nothing physical; it would simply be a device to predict the probability of getting a certain value for a measurement. Such an

function must refer to an ensemble of single entities, but could instead refer to an ensemble of more than one entity.

Before we go on, let me emphasize that whether the wave function described a single entity or multiple entity, it refers to an ensemble. This implies that the entity (or entities) can have different possible configurations, different properties not described by the wave function directly. Each member of the ensemble represented by a given wave function has a different configuration of properties. A given "state" of a system is unique only to the extent that each instance of the state in the ensemble is prepared in a certain defined way; however, that preparation does not fix all the properties. If it did, we would no longer need a statistical theory like quantum mechanics. For example, if the preparation for a coin toss fixed all the properties of each entity involved in the coin toss, then we would always have the same outcome, i.e. not the result of a fair coin toss.

We will, toward the end of the article, add further specification by discussing the natural interpretation[12] (for massive particles) that there are two "entities" statistically described by the wave function in the single particle case: a particle and a guiding wave structure. The specific nature of the particle and the nature of the guiding wave structure are unknown. Two elements can be identified in the guiding wave structure, an organized wave type structure and a stochastic structure;[13] both affect the path of the particle though they are not separately identified in the SE like, for example, the scalar and vector potentials.

Before beginning a detailed analysis of a quantum mechanical system, consider an analogy that brings out the nature of the probabilitistic reference that the square of wave function ($P=|\psi|^2$) provides. An unnamed physicist loses his glasses in his 10-room home. In our analogy,[14] we would like to specify a probabilistic state for where his glasses will be found. Take this state to be specified by a function, $P(room)$ which gives the probability of finding the glasses in a given room *once they are lost*. If where he finds them this time does not affect where he will lose them the next time, then clearly the *probabilistic* state of the system is not changed by his finding them this time, for example, in the living room. Formally, the state remains $P(room)$ even after he found them in the living room. If his future behavior is changed by the find, then the *probabilistic* state is now changed to some new state $\bar{P}(room)$. In general, his finding the glasses, for example, in the living room, does *not* change, does not "*collapse*," the state to: $\bar{P}(living\,room)=1$, for this would reject the very probabilistic model of the state that we mean to assert. Obviously, we do know that finding the glasses in the living room means that they are indeed in the living room, and thus there is 100% chance of finding

---

explanation amounts to what John Bell called a FAPP, a description that is true "*For All Practical Purposes*" but prescinds from talking seriously about the physical reality that one is trying to explore. Leaving aside collapse, what about the contradiction? As mentioned later in the text of the paper, the particle/wave contradiction can be straightforwardly resolved by allowing two entities, a guiding wave structure and a particle.

[12] This interpretation can be debated, especially, when one discusses quantum field theory, but it is reasonable and gives a concreteness that examples always give and that are necessary to make the discussion more clear.

[13] Note this is distinct from standard dBB.

[14] Let me emphasize that this means is similar to, not exactly the same as, quantum mechanics.

them there. However, this is a statement outside the model under discussion which is supposed to give the probabilistic state of finding the glasses *once they are lost*.

Of course, this manifests a key place where this analogy to QM breaks down. In QM, the wave function gives the probabilistic state of the system at every moment. Still, the key point underscored by the analogy stands; the measurement of the system does not "collapse" the system to a non-probabilistic state. It does, in general, evolve the state to a new state. A system prepared in a state $\psi(x,t)$ has a probability $|\psi(x,t)|^2$ of being found between $x$ and $x+dx$. The measurement must involve an interaction with a measuring device characterized by a Hamiltonian, $H_{int}$. The wave function evolves according to the SE: $H_{int}\psi = i\hbar \frac{\partial \psi}{\partial t}$. After the measurement interaction, the state evolves to a new wave function $\bar{\psi}(x,t)$. It may or may not be in an eigenstate of the position operator after the measurement is completed. All that is necessary is that there be a correlation between the reading of the measuring device and the position of the particle before the measurement began. Through this correlation, we can determine where the particle was just before the measurement. Indeed, the measurement process can be summarized formally as follows.[15,16]

Take the state of the measuring device to be given by

(1) $$|\alpha, m\rangle,$$

where:

> the $m$'s are quantum numbers specifying the state of the measuring device that are not directly relevant to our measurement, and
> 
> $\alpha$ represents the number that we get from a macroscopic property that we can read-off the measuring device with our senses directly, such as the readout of a voltmeter. $\alpha$ is set to zero before the measurement.

The measurement interaction evolves the system/measuring-device state by the unitary time evolution operator, $U(t,t_0)$. Starting with the simple case in which the object is in an eigenstate, $|r\rangle$, the measurement then, in general, gives the superposition of states we mentioned:

(2) $$U|r\rangle|0,m\rangle = \sum_{r',m'} u_{r,m}^{r'm'} |r'\rangle |\alpha_r, m'\rangle \equiv |\alpha_r;(r,m)\rangle$$

Note that $\alpha_r$ is in one-to-one correspondence with the relevant eigenstate of the object. In short, reading $\alpha_r$ signifies the object had a value *r* at the time of the measurement. The last equality defines $|\alpha_r;(r,m)\rangle$, which is a helpful notation that labels the complicated sum with the key variables: the *initial* states of the object and measurement device (*r* and *m*) as well as the resulting pointer value, $\alpha_r$.[17] In general, the pre-measured state of the object will be represented by:

---

[15] Reference footnote 4, page 232.
[16] Reference footnote 6, page 245.
[17] We could generalize further by starting with superposition of *m* states. But, this will complicate our notation without substantially affecting the argument, so we leave it aside.

(3) $$|\psi\rangle = \sum_r c_r |r\rangle.$$

And, the resulting state after measurement is written:

(4) $$U|\psi\rangle|0,m\rangle = \sum_r c_r |\alpha_r;(r,m)\rangle.$$

And, again, formally it is clear that the state has not collapsed!

Of course, it is possible that after the measurement, the wave function could, indeed, be left in an eigenstate of the measured observable. For instance, in our position example, this would be a kind of collapsing of the state to $|x\rangle$. Similarly, in our lost-glasses case, it is possible (though not likely) that after finding the glasses in the living room, his behavior is changed so radically that every time later he loses (and thus later finds) his glasses in the living room. This would be a kind of collapsing of the state to $P(living\ room) = 1$.

The point, however, remains that *knowing* where the particle is at a particular moment does not change the statistical nature of the description that the wave function provides. Just because I know that the particle is here at this moment does not mean that I have left my statistical theory. There is simply a new wave function to describe the system after the measurement is complete, which is, in turn, simply a representation of the new statistical state that the system has entered. Remember, by statistical state we mean an ensemble of systems prepared in the same way.

## Stern Gerlach Experiment

To understand measurement concretely, consider the experiment that starts with a neutral spin ½ particle in the initial state:

(5) $$|n\rangle_0 = |\rightarrow\rangle|\psi\rangle = \frac{1}{\sqrt{2}}\left(|\uparrow\rangle|\psi\rangle + |\downarrow\rangle|\psi\rangle\right).$$

Here $\psi$ describes the spatial part of the state of the particle, and the arrows indicate its spin state. In our case, we take $\psi$ to be a plane wave moving with some momentum $p = \hbar k$. $|n\rangle_0$ describes the state of an ensemble of one (neutral) particle systems prepared in the same way. In particular, it describes a system prepared in such a way that an arbitrarily accurate measurement of momentum of any member of the ensemble of possible systems will yield the value *p*. Now, sending such a particle through an up/down Stern Gerlach device (SG) will evolve the state so that at time $t_1$ it becomes:

(6) $$|n\rangle_1 = \frac{1}{\sqrt{2}}\left(|\uparrow\rangle|\psi_\uparrow\rangle + |\downarrow\rangle|\psi_\downarrow\rangle\right)|SG\rangle$$

The spatial part $\psi_\uparrow$ ($\psi_\downarrow$) is a wave function redirected along an upward (downward) path. Along the upward path we will find only upward particles. We have spatially separated the particles, but the ensemble of particles created in this way share a common spin state structure as indicated by the superposition distinguished only by spatial path. This can be made evident in the following way.

Redirect these paths, with appropriate particle analogs of optics (e.g., particles lenses and mirrors) to rejoin so that: $|\uparrow\rangle|\psi_\uparrow\rangle|SG\rangle \to |\uparrow\rangle|\psi\rangle|SG'\rangle$, $|\downarrow\rangle|\psi_\downarrow\rangle|SG\rangle \to |\downarrow\rangle|\psi\rangle|SG'\rangle$, and so we get at time $t_2$:[18]

$$|n\rangle_2 = \frac{1}{\sqrt{2}}\left(|\uparrow\rangle|\psi\rangle + |\downarrow\rangle|\psi\rangle\right)|SG'\rangle \quad (7)$$

This is fundamentally the same particle state as before the SG![19]

## Wigner's Friend
### *Ensemble Interpretation*

Now, consider the so-called Wigner's friend paradox.[38] Wigner's friend, $F$, sends a neutral spin ½ particle towards an up/down Stern Gerlach device (SG) as described above. Then, $F$ shoots a light beam at the up output region. If he sees the light reflect back then his brain registers (in a complex way) this fact and concomitant to this physical change of state of his brain, $F$ comes to know that the particle is spin up. This last fact, i.e., $F$'s new knowledge, in so far as it is knowledge (i.e. distinct from the physical change in the brain[20]), is not described by the SE.[21] There is no separate physical interaction, and no Hamiltonian to evolve the state further than the state change of his brain. The equation allows for no such further change in state. However, the change in state of the light and the brain arise from physical interactions and *are* (as far as we know)[22] described by the SE through the proper interaction Hamiltonians. The subsequent evolution due to these interactions gives:[23]

---

[18] If SG keeps a record in some way of which way the particle goes then we would have:
$\left(|\uparrow,\psi_\uparrow\rangle|SG\rangle_\uparrow + |\downarrow,\psi_\downarrow\rangle|SG\rangle_\downarrow\right)/\sqrt{2}$.

[19] More accurately, the particle state is the same state as before except translated along, for example, the x-axis. Of course, the full guiding wave structure is different because one must include the change in SG state.

[20] Clearly, the physical state of the brain when one sees yellow is not the same as the act of seeing yellow. See Rizzi, *Science Before Science: A Guide to Thinking in the 21st Century* (SBS) (IAP Press, Baton Rouge, 2004).

[21] One should not expect SE to describe cognitive action. One should not assume what one does not have evidence for. SE describes an ensemble of systems that have been, as far as can be, prepared in the same way. This preparation does not succeed in making them exactly the same. Because of this, our equation cannot describe an individual system but only an ensemble of systems. In this way, it is like the case of a well shuffled deck of cards. I cannot predict what card I will draw. However, I can predict what the chances of getting a given card is. After I draw a card (say without looking at it), the probability distribution has changed, because now I have pulled a card out and the deck now belongs to a different ensemble of well shuffled cards; it belongs to an ensemble of well shuffled decks that are missing the particular card that I have drawn out. This new ensemble has a new probability distribution, a new $P(x_i)$, where $x_i \in \{A_{club}, K_{club}, Q_{club}...A_{spade}, K_{spade}...A_{heart}, K_{heart}...A_{diamond}, K_{diamond}...\}$ associated with it. Suppose, now, I look at the card I have drawn to find that it is the king of clubs. I have not now changed the probability distribution just by knowing what it is! I do not collapse the probability distribution of the ensemble, by gaining knowledge of the card. The act of withdrawing the card changed the system in a fixed way, not my act of knowing that I drew out a card. My *knowing* what card I drew does not change the physical state of the cards. This is so despite the fact that it does allow me to know to what ensemble my new depleted deck now corresponds.

(8) $$n_2 = \frac{1}{\sqrt{2}}\left(|\uparrow\rangle|\psi_\uparrow\rangle|light\rangle|F\rangle_\uparrow + |\downarrow\rangle|\psi_\downarrow\rangle|no-light\rangle|F\rangle_\downarrow\right)|SG''\rangle$$

The *F* kets here describe the state of the Friend's brain and thus the state of his knowledge. Note that before this last measurement, the state of the light and *F*'s brain were not entangled with the particle so, like the rest of the universe, we did not reference their state in equations (5) and (6).

Where is the paradox you ask? We have to introduce Wigner himself. We must ask Wigner what he thinks. Suppose he is told what his friend will do. Then he can calculate the state and he will agree with his friend that it is given by equation (8). Still, no paradox! Indeed, there is no problem if one understands quantum mechanics in this natural way. However, there is, as we will see, a problem when using the orthodox interpretation.

*Copenhagen as Contrasted with Ensemble Interpretation*

The orthodox interpretation introduces an extra wave function evolution (the previously mentioned *R* (reduction) process) that is not described by the SE. This reduction mechanism collapses the wave function to the eigenstate of the measured property corresponding to the value one measures (i.e., the value one now knows) for the particular case. This approach proceeds as if, at least once the measurement is made, the wave function describes a single system rather than an ensemble.

If the wave function described the single system under analysis, once the value of the given property is known, one would have to change the wave function to correspond to it. In this reasoning if one measures a particle to be at position *x*, it must be in an eigenstate of position $|x\rangle$. By treating the state in this way, one acts as if the measurement (knowledge) in some way allows one to enter a non-probabilistic model. Again, this is false, as the state described by SE describes ensembles, as the SE gives only probabilities.

By measuring the particle, I have only entangled my brain and the measuring device with it, not reduced it to a simple eigenstate. It is true that I can use this knowledge to create a new system that approximately is in an eigenstate $|x\rangle$. However, this does not belie the fact that in another experiment prepared in the same state --i.e., a different member of the ensemble defined by the initial wave function, $\psi_{initial}(x,t)$, will, in general, detect the particle somewhere else. Nor does it belie the fact that the wave function, in general, goes multiple places after a measurement, and I am, through my approximation, ignoring that larger wave function. The system is only in some limited sense now described by $|x\rangle$.

---

[22] For more on the mind, brain and physical theory, see Chapter 5, *The Science Before Science*, referenced in footnote 20.

[23] We now have a many particle QM system. The ensemble interpretation handles these systems straightforwardly. This includes issues related to Bell's theorem. As is widely known, Bell's theorem, via several assumptions that seem verified by experiment (cf. ref 6), shows that there are superluminal connections between particles. Indeed, quantum mechanical treatments would use configuration space (thereby imposing non-locality) to formally treat hidden variables when these connects are discussed. This occurs, for example, in the de Broglie-Bohm formalism.

To see this, take the example of plane wave associated with a neutron incident on a screen with small hole in it. Imagine that the wave function describes, as mentioned earlier, a complex (stochastic) guiding wave structure acting on a particle. In this case, a given member of the ensemble described by this plane wave has a definite state of this structure as well as a neutron in a definite location.[24] Because of this, in some experiments the neutron will go through the hole, and, in others, it will not. Until there is further interaction, the wave function will simply evolve according to the interaction with the screen as shown here in Figure 1a .

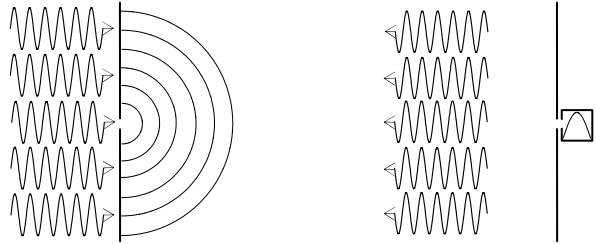

Figure 1. a) neutron incident on a screen with a single hole in it. The incoming neutron state is represented by a rightward traveling plane wave of finite extent.  b) same experimental setup after insertion of a box that captures the neutrons that escape through the hole. Here we assume that, upon impact, the wave function reflects off the screen and thus becomes a leftward traveling plane wave of finite extent. The curve in the box represents the ground state wave function of the neutron in the box, which we assume is the only state the neutron can populate. The wave function is shown after the reflection of the finite length wave train.

Note, as shown in Figure 1b, the measurement of the particle coming through the hole does not collapse the full state to $|x\rangle$, but simply changes the evolution of the ensemble. It only changes it in some ways because, for example, some members of the ensemble still will not pass through the hole at $x$. To get our approximate eigenstate of position, we must ignore all those that reflect.

*The Paradox itself: Wigner's Friend in Copenhagen Interpretation*

Let's see now why a paradox arises in the orthodox interpretation; in fact, a contradiction arises not just a paradox. If one assumes that a collapse happens, then *F*'s measurement would, instead of giving state (8), give:

(9) $\quad |\uparrow\rangle|\psi_\uparrow\rangle|light\rangle|F_\uparrow\rangle|SG''\rangle \quad or \quad |\downarrow\rangle|\psi_\downarrow\rangle|no-light\rangle|F_\downarrow\rangle|SG''\rangle$

or written more simply: $\quad |\uparrow\rangle \quad or \quad |\downarrow\rangle$

(which can be written as *mixed* state:[25]

$\frac{1}{\sqrt{2}}\left(|\uparrow\rangle\langle\uparrow|+|\downarrow\rangle\langle\downarrow|\right)$ , notice absence of cross terms

---

[24] Note that there will be commonality between the different instances of this structure in each member of the ensemble because of the common preparation reflected in the fact that the ensemble is described by the same wave function.

[25] The following is, indeed, a matrix but, in this usage, the matrix represents a statistical state.

which signals that this is very different from the pure state $(|\uparrow\rangle+|\downarrow\rangle)(\langle\uparrow|+\langle\downarrow|)$ )

Whether we think the state is given by (9) or by (8) drastically changes our interpretation of what happens; indeed, it leads to the Wigner's friend paradox.

To see this, recall Wigner's view of his friend's experiment. He is outside the friend's lab. Assume that his friend was given instructions to do what he did at a certain time. Then, using the ensemble interpretation, as we have noted, everything goes through fine. Wigner and the friend will both describe the state by (8). Wigner will know everything that his friend does *except* the value he measures for the individual member of the ensemble. By contrast, if we suppose collapse in the manner described above, then Wigner says the state is given by (8), whereas his friend has collapsed the wave function by his knowing the result and describes it by the mixed state given in (9). Hence, the paradox, which is actually a simple contradiction. The system cannot be in both states as they are (obviously) different.

*More Depth on the Ensemble Understanding of Wigner's View*

To further understand the situation as revealed by the ensemble interpretation, suppose that the experiment is repeated many times, then $F$ will have knowledge of which members of that subset of the potentially infinite ensemble represented by (8) are up and which are down. However, the state will still be described by (8). Unlike Wigner, $F$, can make a new ensemble that just consists of those in the up particles. Say, he has a whole line of these experiments in his lab. He can have a box at the end of the up trajectories that captures the particle.[26] Then, he can collect all the boxes in the line and put them in one place. Considering only those boxes, he will then have a sub ensemble of the original one that can be considered, in some approximation, as new ensemble of the particle in the up state. Wigner could not do this because he does not know which boxes in the original line have up particles in them. More generally, $F$ and Wigner are in a different states with respect to the particle, so we would expect asymmetry between them.

Going back to having light illuminate the up particle trajectory, we can make this point more manifest. We can, in contrast to (8), explicitly include Wigner, in which case, the state (after the measurement) is:

(10) $$\frac{1}{\sqrt{2}}\left(|\uparrow\rangle|\psi_\uparrow\rangle|light\rangle|F\rangle_\uparrow+|\downarrow\rangle|\psi_\downarrow\rangle|no-light\rangle|F\rangle_\downarrow\right)|SG''\rangle|W\rangle.$$

Notice that Wigner's brain ($W$) is shown in a state that is not entangled with the particle. This is a representation of the fact that, at the end of any given experiment, Wigner, unlike his friend, does not know what state the particle is in. The asymmetry between $F$ and $W$ is evident. $W$ and $F$ would both calculate that the state is given by (10). They both know, from this state, that $F$ will know the results of each measurement, while $W$ will not. Again, note well that *knowledge* per se of the result of a measurement does not itself change the state; it is the *physical interactions* associated with the measurement that does.

## More on Stern Gerlach Experiment

To better understand how the statistical state (i.e. the ensemble of possibilities represented by the wave function) changes when one measures[27] which direction the

---

[26] This is not described by our current expression for the state but can easily be accounted for.

[27] By "measuring the state of X," we mean the act of someone reading a measurement device whose reading (pointer state) has been determined by the physical state of X and that can affect the senses in such

particle goes in an SG experiment, consider the following. If we try to recombine the two halves as we did in the first section, we would obviously get (8), instead of (6), which leaving aside the state of the SG is:

(11) $$|system\rangle \equiv \frac{1}{\sqrt{2}}\left(|\uparrow\rangle|\psi_\uparrow\rangle|light\rangle|F_\uparrow\rangle + |\downarrow\rangle|\psi_\downarrow\rangle|no-light\rangle|F_\downarrow\rangle\right)$$

Introducing the more general density matrix (state operator) notation, this can be written:

$$|system\rangle \equiv \frac{1}{\sqrt{2}}\left(|\uparrow\rangle|\psi_\uparrow\rangle|E_\uparrow\rangle + |\downarrow\rangle|\psi_\downarrow\rangle|E_\downarrow\rangle\right)\left(\langle\uparrow|\langle\psi_\uparrow|\langle E_\uparrow| + \langle\downarrow|\langle\psi_\downarrow|\langle E_\downarrow|\right),$$

where $E$ stands for the environment of the particle.

Now, if we ignore the environment, i.e., the light and the state of the brain of $F$ by tracing over them, we get the state of the particle to be:

(12) $$tr_E|system\rangle\langle system| = \sum_1^2 \langle E_i|system\rangle\langle system|E_i\rangle$$

$$= \langle l, F_\uparrow|system\rangle\langle system|l, F_\uparrow\rangle + \langle no-l, F_\downarrow|system\rangle\langle system|no-l, F_\downarrow\rangle$$

$$= \frac{1}{2}\left(|\uparrow\rangle|\psi_\uparrow\rangle\langle\uparrow|\langle\psi_\uparrow| + |\downarrow\rangle|\psi_\downarrow\rangle\langle\downarrow|\langle\psi_\downarrow|\right)$$

So, we have a mixed state, which represents an ensemble whose elements are radically different from that of the pure state: $\frac{1}{2}(|\uparrow\rangle|\psi_\uparrow\rangle + |\downarrow\rangle|\psi_\downarrow\rangle)(\langle\uparrow|\langle\psi_\uparrow| + \langle\downarrow|\langle\psi_\downarrow|)$. Each member of the pure state has a coherence that the members of the mixed state do not have. Or, said another way, the particle state given by equation (11) has a coherence with the environment that disrupts its simple spin coherence present in the state represented by (5).

This leads us to discuss the previously mentioned guiding wave structure. The presence of coherence implies the guiding wave structure associated with this particle state can interfere to create patterns of behavior that otherwise could not occur. Because that structure is stochastic the patterns will have a certain statistical fuzziness to them. The environment breaks the ability of the wave structure to interfere. An analogy is helpful here. Consider two point sources emitting E&M radiation in phase with each other. One might think the waves thus generated would necessarily interfere; however, they *cannot* interfere if one source emits horizontally polarized light and the other emits vertically polarized light. This situation is analogous to the guiding wave structure that would result from the interaction between two point sources of neutrons described by in-phase wave functions. An active neutron detector placed near one of the two sources would become entangled with that source and thus, to some degree, prevent the wave structure from that source from "interfering" with that of the other source. This is clear with the SG experiment, for as we saw if someone puts a detector in the up stream, he destroys the ability of the wave structure to be recombined, to "interfere" to reproduce the original state like can be done if no such detector entanglement occurs.

---

a way to change the state of the brain. Recall that while the physical states of the senses and brain are necessary for your knowledge of the state of X, they are not identical with that knowledge.

## Free Particle States and the Schrödinger Cat in a Fuller Interpretation

To bring out the meaning of the ensemble interpretation in a more concrete way, we now turn to direct consideration of the guiding wave structure associated with a single particle system, especially through analysis of two important systems.

The guiding wave structure that we introduce here has an affinity with another interpretation of quantum mechanics, the de Broglie-Bohm interpretation (dBB); yet, it is very different. DBB takes the particle to be *precisely* guided by a wave represented by the wave function.[28] Our discussion does not depend on this assumption. We do not assume that QM gives the exact trajectories; indeed, all our analysis implies otherwise. Why would rearranging a statistical equation (the SE) suddenly give exact trajectories?[29] We only note that QM indicates the presence of some stochastic wave structure whose exact nature is unknown, but which, in some way, influences how and where the particle moves. However, dBB is not irrelevant. After all, it is a formal rearrangement of the SE, so one expects it contains no more information but, properly handled, no less either. Indeed, dBB gives, for example, the average velocity (of members of the ensemble in which the particle is at position $x$)[30] under this "stochastic wave structure" influence.

### *Free States*

Now, take the simple free particle state $|p\rangle$, which in the position basis is: $\langle x|p\rangle = e^{i\frac{px}{\hbar}}$. In terms of the full time-dependent wave function, the state is written: $\psi(x,t) = e^{i\left(\frac{px}{\hbar} - \omega t\right)}$, which is a plane wave of wave vector $k = p/\hbar$, and where $\omega = \frac{1}{\hbar}\frac{p^2}{2m}$. This state consists of an ensemble of particles, all of which under an arbitrarily accurate measurement of momentum yield the value $p$. According to the natural interpretation adopted here, this wave aspect corresponds to an element of the guiding wave structure present in *every* member of the ensemble. In addition, each member of the ensemble represented by this $\psi$ has further activity that *varies* from one member to the next. This second element of the guiding wave structure is stochastic and results from various (unknown) agents.[31] Each member of the ensemble can be viewed as containing a single particle, which can be *measured* to be moving precisely at a velocity, $p/m$,[32] accompanied by this just-described guiding wave structure.

---

[28] The guiding of the particle by the wave is not simply read off the wave function but is described via differential equations derived from the SE by the substitution $\psi = \text{Re}^{iS}$.

[29] See ref. 6 for more discussion on dBB. There (on page 323) it is pointed out that it would not make any sense to suppose that from the Maxwell thermodynamic relations, which relate things like entropy and temperature, one can deduce the trajectories of the particles that make up a thermodynamic system.

[30] See ref 6, page 68.

[31] Stochastic here refers to not being correlated with any known physical system.

[32] Said another way, the particles can be said to be moving at a constant *average* velocity in this interpretation. Currently understood measurement interactions (and perhaps measurement interactions in principle) may, for example, effectively integrate over a finite (though very small) time scale because they have an inherent limit in their response time. If this were an *in principle* limit, QM would then also be describing, though indirectly, this aspect of nature.

The next most complex state is represented by constructing a superposition of such plane waves. Consider, for example, the Gaussian wave function given by:

$$\Psi(x,t) = \frac{1}{(2\pi)^{1/4} \sigma(t)/\sqrt{\sigma_0}} e^{-\frac{x^2}{4\sigma(t)^2}} \tag{13}$$

$$\text{where } \sigma(t) = \sigma_0 \sqrt{1 + i\frac{t}{\tau_0}}, \text{ with } \tau_0 \equiv \frac{2m\sigma_0^2}{\hbar}.$$

This state corresponds to an ensemble each member of which has a particle moving at a velocity[33] determined by the probability weighting corresponding to the plane wave decomposition. And, each has a guiding wave structure, the key element of which is described by the weighted sum of these plane waves that gives the Gaussian wave function. Take the case of the state at $t=0$, the Fourier decomposition is:

$$f(k) = \frac{1}{(2\pi)^{1/4}\sqrt{\sigma_0}} \int_{-\infty}^{\infty} e^{-\frac{x^2}{4\sigma_0^2}} e^{-ikx'} dx' = (2^3 \pi)^{1/4} \sqrt{\sigma} e^{-k^2\sigma_0^2} \propto e^{-k^2\sigma_0^2} \tag{14}$$

Thus, recalling that $v = \hbar k/m$, we see that the ensemble of particles represented by the wave function in equation (13) has a Gaussian distribution of velocities centered on zero with a standard deviation of $\sigma_v = \frac{1}{m}\frac{\hbar}{\sigma_0}$. Each member of the ensemble is also associated with a wave structure distributed in frequency according to equation (14).

### *Schrödinger's Cat Paradox*

Lastly, given this simple understanding of the wave function and the SE, we can clear the Schrödinger's cat to definitively have life or death. Schrödinger introduced his paradox to manifest the deep problem with the Copenhagen interpretation. In the Copenhagen interpretation, a cat in the superposition state

$$|cat-live\rangle + |cat-dead\rangle, \tag{15}$$

is neither dead nor alive until it is measured to be in one or the other! One can see why Einstein asked if Copenhagen supporters really thought the moon wasn't there until they looked at it?[34,35,36,37,38]

---

[33] By "velocity," I here mean the value obtained if a given member of the ensemble were measured.

[34] A. Pais said "that during one walk Einstein suddenly stopped, turned to me and asked whether I really believed that the moon exists only when I look at it." (footnote 35) The authority of Wigner is historically important for the idea that consciousness "collapses the wave function." Eugene Wigner explained that collapse requires consciousness to terminate the otherwise infinite regress of physical causes and actually have the measurement we see (see footnote 36). (Without consciousness, which would be an *R* process, each physical measurement would require something to collapse it and the system would never actually be in any definite state.) Later in life, he said this argument led to solipsism and thus was a kind of *ratio ad absurdum* that leads to the conclusion that quantum mechanics is incomplete in the radical sense of not making any sense as it currently stands (see footnote 37). To not lead to absurdity, QM requires, according to him, a physical (not cognitive) collapse mechanism that it currently does not have. This conclusion arises through his acceptance of the orthodox ("collapse") interpretation of quantum mechanics. Note well that the incompleteness with which Wigner is here concerned is radically different from the incompleteness present in any statistical theory. A statistical theory is not a formally complete theory because it can only

This understanding can arise when one assumes that $\psi$ represents a single entity, not an ensemble each of whose members has multiple entities. In this case, one takes $|cat-live\rangle$ to mean the cat in front of me is alive and $|cat-dead\rangle$ to mean it is dead.[39] Then, what do I say of my cat in the superposition state (15)? It must be both, waiting to be one or the other! That doesn't mean much if anything, but that seems to be as far as that interpretation is willing to go in explaining the meaning of the state.[40]

This dead/alive cat is easily resolved in the ensemble interpretation. The ket represents an ensemble of possibilities.[41,42,43] This means that in the superposition state (15) consists of an ensemble some members of which have a dead cat and some members have an alive cat. And, in the filled out view, each member has a guiding wave structure one element of which is the same as any other member. However, because there is also a stochastic aspect of that structure, each member's guiding wave structure is different in some way from the next. As in the other example, the measurement interaction entangles the cat with the *detector* and the experimenter's *brain* (call these later two "DB") but does not change the probabilistic nature of the theory.[44] Formally we write:

(16) $$|cat-live\rangle|DB_{cl}\rangle+|cat-dead\rangle|DB_{cd}\rangle$$

Finding the cat alive does not collapse the state to $|cat-live\rangle|DB_{cl}\rangle$. There is still a second branch. This does not mean that there remains a possibility that the cat is dead (and, a fortiori, it does not mean that, in addition to being alive, the cat is *also* dead). It

---

tell probable outcomes and thus is missing a reference to part of the causal structure of the reality under study.

[35] A. Pais, Rev. Mod. Phys. **51**, 863 (1979) pg 907.

[36] Wigner says that the "…argument for the difference in the roles inanimate observation tools and observers with consciousness--hence for the violation of physical laws where consciousness plays a role--is entirely cogent so long as one accepts the tenets of orthodox quantum mechanics in all their consequences." (see Mehra, ref. 38)

[37] J. Mehra,. (ed.) *The Collected Works of Eugene Paul Wigner,* Part B, Historical Philosophical and Socio-Political Papers. Vol. 6: Philosophical Reflections and Syntheses (Springer, Berlin, 1995) pg 593.

[38] Wigner, *Remarks on the Mind-Body Question* in I. J. Good, (ed.) *The Scientist Speculates* (London: William Heinemann, Ltd., 1961; New York: Basic Books, Inc., 1962), ch. 13, pg 181. See also Ballentine, *Foundations of Physics* **9**, 783 (2019)

[39] In the many worlds interpretation, one would say, for our cat case, there are two universes, one in which the cat is alive and one in which he is dead. The universes split from each other when a measurement is done. In this way, the wave function describes multiple entities that exist at a given moment, but they are, in some sense, copies of themselves. There is no probabilistic ensemble. The probabilistic ensemble has been reified from possible outcomes into concurrently-existing outcomes. It is as if one were to decide that, in a dice roll, all the different landing possibilities already exist, but in different universes! The *single* particle wave function thus represents many universes (instead of a probabilistic ensemble), but one assumes that only *one* entity exists in each universe, for example, a spin-up particle. This single entity assumption is also unlike our concrete interpretation, which has, for example, a guiding wave structure with a wave aspect and a stochastic aspect as well as a particle.

[40] More complex discussions attempting to explain this don't seem to add much to the bald statement.

[41] PBR (reference 42) has shown that each ket specifies a unique "state," i.e. each member of the ensemble represented by a given ket is a unique (possible) physical state.

[42] M. F. Pusey, J. Barrett, T. Rudolph, *On the reality of the quantum state*, Nat. Phys. **8**, 475–478 (2012)

[43] A. Rizzi, *Does the PBR Theorem Rule out a Statistical Understanding of QM?* Found. of Phys., **48** (12), 1770-1793.

[44] See page 1785 of ref. 43.

means that, in a similarly prepared experiment, the experimenter might find the next cat to be dead!

## Conclusion

We have seen that the ensemble interpretation is a natural interpretation of the SE that provides an accessible, and unstrained understanding of measurement in quantum mechanics. Students access to quantum mechanics will be greatly aided by presenting QM in this way. The discussion brings out three central points that have been neglected by most authors, at least by lack of emphasis. *First*, quantum mechanics is a statistical theory; it predicts probabilities not single outcomes. As such it demands consideration of ensembles of measurements, not single measurements; hence the primacy of *ensembles*. *Second*, it does not require the assumption that $\psi$ represents ensembles each with only a **single** entity. Contradictions can even result by forcing this single entity assumption, for example forcing one to posit an entity that is both a particle and wave and the same time.[11] *Thirdly*, in understanding the measurement process, one needs to avoid confusing the physical state of the brain and the act of cognition associated with it. If this distinction is not clearly made, one can end in making the existence of the external universe depend on our own thought about it! In summary, quantum mechanics is about ensembles each member of which has multiple *physical* (not cognitional) entities. Once the three points are recognized, long standing conundrums such as the Schrödinger cat and the Wigner friend's paradoxes are straightforwardly resolved in the ensemble interpretation. Lastly, the further developed ensemble interpretation given here recognizes the presence of wave structure, stochastic structure and (likely) a particle, and gives a less abstract view of the nature of the ensembles involved in QM. Furthermore, it is able to wield the strengths of the de Broglie-Bohm interpretation without binding one to its ad hoc assumption that QM can give formulae for exact trajectories.

It may be shown in the future that quantum mechanics is incorrect at some level (indeed, almost certainly will be, given its manifest incongruity with general relativity) and thus will be supplanted by a future theory that includes present quantum mechanics as a limiting case. However, until then, and probably after then, quantum mechanics has no collapse mechanism but is a statistical theory in which the wave function evolves under a linear unitary transformation in time. So, it is natural and useful to treat it as such.